\documentclass[showpacs,superscriptaddress,preprintnumbers,amsmath,twoside,amsfonts,twocolumn,amssymb,pra]{revtex4-1}
\usepackage{geometry}
\usepackage[utf8x]{inputenc}
\usepackage{amsmath}
\usepackage{amsfonts}
\usepackage{amssymb}
\usepackage[dvipsone]{graphicx}
\usepackage{nicefrac}

\def \be {\begin{equation}}
\def \being {\begin{equation*}}
\def \en {\end{equation}}
\def \ening {\end{equation*}}
\def \beray {\begin{array}{ccl}}
\def \enray {\end{array}}
\def \d {^\dagger}
\def \a {\alpha}
\def \b {\beta}
\def \g {\gamma}
\def \de {\delta}

\begin{document}

\title{Laser Plasma Interaction and Non-classical Properties of Radiation Field}
\author{Aabhaas Vineet Mallik}
\email{aabhaas.iiser@gmail.com}
\affiliation{Indian Institute of Science Education and
Research-Kolkata, Mohanpur 741252, India}
\author{Pratyay Ghosh}
\affiliation{Indian Institute of Technology-Bombay, Mumbai, India}
\author{Ananda Dasgupta}
\email{adg@iiserkol.ac.in}
\affiliation{Indian Institute of Science Education and
Research-Kolkata, Mohanpur 741252, India}

\date{\today}

\begin{abstract}
We show by explicit calculations that non-classical states of the radiation field can be produced by allowing short term interaction between a coherent state of the radiation field with plasma. Whereas, long term interaction, which thermalizes the radiation field, can produce non-classical states of the radiation field only at sufficiently small temperatures. A measure of $k^{\text{th}}$ order squeezing, stricter than the one proposed by Zhang et al \cite{Zhang}, is used to check the emergence of squeezing. It is also shown that photons in the considered thermalized field would follow super-Poissonian statistics.
\end{abstract}

\maketitle

\section{Introduction} Most of the light sources around us are classical in nature; in the sense that all the properties of the emitted radiation can be explained by assuming that the emission process in the source is actually some classical stochastic process. To put in more pedagogical terms, a radiation field is said to be `classical' if the Glauber-Sudarshan $\mathcal{P}$-function \cite{Sud}\cite{Glaub1}\cite{Glaub3} corresponding to the state of the radiation field, is positive definite.

It turns out that there can be states of the radiation field for which the $\mathcal{P}$-function is either highly singular (for example, containing derivatives of the Dirac delta function) or is not positive in the whole parameter space (sometimes also referred to as the phase space). Properties of such radiation fields cannot be explained by modeling the emission process by any classical stochastic process. It is with these light sources that one can exploit the full power of the quantum nature of the electromagnetic field. Some of the properties which can only be exhibited by `non-classical' radiation are antibunching, sub-Poissonian distribution of the number of photons in the field (sub-Poissonian statistics) and squeezing. 

One way to produce non-classical radiation is via the interaction of a coherent radiation (easily obtained from a laser operating well over threshold) with free electrons \cite{Ben}. In this work we calculate the short term and long term\footnote{Short and long compared to the decoherence time scale of the system.} non-classical effects of interaction, on the incoming coherent radiation, with the free electron gas. We explicitly show that short term interaction can be used to produce non-classical states of radiation field exhibiting squeezing. And contrary to expectation, even long term interaction, which arguably decoheres the radiation field, can be used to produce non-classical radiation at sufficiently small temperatures.

\section{Field Hamiltonian in presence of plasma}
Ben-Aryeh and Mann \cite{Ben} have shown that under certain approximations the Hamiltonian for the pair of modes, $\mathbf{k},\lambda$ and $-\mathbf{k},\lambda$ of the radiation field interacting with slow moving and sparse `free' electrons (which, from here on, we will refer to as plasma) in a volume having dimensions larger than $|\mathbf{k}|$ can be written as
\be
H_{\mathbf{k},\lambda} = H_+ + H_-
\en
where
\be
H_+ = \omega a\d _+ a_+ + \Omega_1 \left[ a\d _+ a\d_+ + a_+ a_+ \right] + \Omega_2 a\d_+ + \Omega_2^* a_+
\en
\begin{eqnarray}
a_+ = \left( a_{\mathbf{k},\lambda} + a_{-\mathbf{k},\lambda}\right) / \sqrt{2}\\ a_- = \left( a_{\mathbf{k},\lambda} - a_{-\mathbf{k},\lambda}\right) / \sqrt{2}.
\end{eqnarray}

$H_-$ is obtained from $H_+$ by replacing all the $+$ in the subscript by $-$. $\omega$ and $\Omega _1$ are real constants related to $\omega_{k} = ck$ and the density of free electrons in the plasma, and $\Omega _2$ is a complex number. $a_+$ and $a_-$ are bosonic annihilation operators for two uncoupled modes of standing waves.

For our calculations we start with a more general mode pair Hamiltonian\footnote{We drop the +/- in the subscript for brevity.},
\be
H = \frac{\omega}{2} \left( a \d a + \frac{1}{2} \right) + \frac{\Omega _1}{2} a ^{\dagger 2}  + \frac{\Omega _1^*}{2} a ^2 + \Omega_2 a \d + \Omega _2^* a
\en
$\omega$ is a real constant, whereas, $\Omega _1$ and $\Omega _2$ are constant complex numbers. One can easily identify $\nicefrac{1}{2} \left( a \d a + \nicefrac{1}{2} \right)$, $ \nicefrac{a ^{\dagger 2}}{2}$ and $\nicefrac{a^2}{2}$ in eqn(5) with the elements of the SU(1,1) algebra, $K_3$, $K_+$ and $K_-$, respectively (harmonic oscillator realization of SU(1,1) algebra). $K_3$, $K_+$, $K_-$, $a$ and $a \d$ together form the elements of the {\it double photon algebra} \cite{ADG2}. Hence,
\be
H = \omega K_3 + \Omega_1 K_+ + \Omega_1^* K_- + \Omega _2 a \d + \Omega_2^* a.
\en

This is the most general Hermitian operator that can be constructed from the elements of the double photon algebra, modulo a real constant. Eqn(6) can be rewritten as
\be
H = \frac{1}{2} \tilde{a} \d A \tilde{a} + \tilde{\Omega} \d \tilde{a}
\en
where
\be
\tilde{a} = \left( \begin{matrix} a \\ a \d \end{matrix} \right)
\en
\be
\tilde{\Omega} = \left( \begin{matrix} \Omega _2 \\ \Omega_2^* \end{matrix} \right)\en
\be
A = \left( \begin{matrix} \nicefrac{\omega}{2} & \Omega _1 \\
				   \Omega_1^* & \nicefrac{\omega}{2} \end{matrix} \right).
\en

$H$ can be simplified to an element of the SU(1,1) algebra (apart from a c-number) by the unitary transformation described below. For
\be
\tilde{\alpha} = \left( \begin{matrix} \alpha \\ \alpha ^* \end{matrix} \right) \equiv A^{-1} \tilde{\Omega}
\en
\be
H' \equiv D(\alpha) H D \d (\alpha) = P - c
\en
where, $D(\alpha)$ is the displacement operator defined as
\be
D(\alpha) = \exp\left( \alpha a \d - \alpha^* a \right)
\en
\be
P = \frac{1}{2} \tilde{a} \d A \tilde{a} = \omega K_3 + \Omega _1 K_+ + \Omega_1^* K_-\\
\en
and
\be
c = \tilde{\Omega }\d A^{-1} \tilde{\Omega}.
\en

\section{Unitary Time evolution}
The state of the radiation field would undergo unitary time evolution for interaction time small compared to the decoherence or relaxation time scale of the system. For such small interaction times the unitary time evolution operator for the radiation field is given by $(\hbar = 1)$
\be
\begin{array}{ccl}
U(t) & = & \exp\left( -i H t \right) \\
 & = & D \d \left( \alpha \right) \exp \left( -i H' t \right) D(\alpha) \\
 & = & e^{ict} D \d (\alpha) \exp(-iPt)D \left( \alpha \right). \end{array}
\en

The initial state of the system under consideration is a coherent state. If $\Omega _1 = 0$, a coherent state would remain coherent at all times and no non-classical properties can be observed. For our system $\Omega_1 \neq 0$, but owing to the small density of free electrons in the plasma, $|\Omega_1 | \ll \omega$. Nevertheless, in the sections to follow we will show that even a small $\Omega _1$ leads to evolution of non-classical properties in the radiation field.

\subsection{The Disentanglement Formula}\label{dis}
To deal with coherent states it is best to write the SU(1,1) group element in eqn(16) in the following disentangled form,
\be
e^{-iPt} = e^{\beta K_+} e^{\gamma K_3} e^{\delta K_-} \equiv V.
\en
That we can always do this, ie, such $\beta$, $\gamma$ and $\delta$ would always exist is guaranteed by the Wei-Norman decoupling theorem \cite{WeiN}. To find $\beta$, $\gamma$ and $\delta$ in terms of $\omega$ and $\Omega_1$ we consider the following relations.
\be
\begin{array}{ccl} e^{-iPt} \tilde{a} e^{iPt} & = & e^{-it(ad\ P)} \tilde{a}\\
  & = & e^{-it\tilde{P}}\tilde{a} \end{array}
\en
where the super operator $(ad\ P)$ is defined recursively as
\be
(ad\ P)^n A = [P,A]_n = [P,[P,A]_{n-1}]
\en
and
\be
\tilde{P} = \left( \begin{matrix}-\nicefrac{\omega}{2} & -\Omega_1 \\
\Omega _1^* & \nicefrac{\omega}{2} \end{matrix} \right).
\en
One can show that
\be
e^{-it\tilde{P}} = \left( \begin{matrix} \cos \phi t +i \frac{\omega}{2\phi}\sin \phi t & i\frac{\Omega_1}{\phi}\sin \phi t \\
-i\frac{\Omega_1^*}{\phi}\sin \phi t & \cos \phi t - i \frac{\omega}{2\phi}\sin \phi t\end{matrix}\right)
\en
where $\phi = \sqrt{\nicefrac{\omega^2}{4} - |\Omega_1|^2}$. Using eqn(21) in eqn(18) and comparing it with
\be
VaV^{-1} = e^{\nicefrac{-\gamma}{2}}( a - \beta a \d )
\en
\be
Va\d V^{-1} = (e^{\nicefrac{\gamma}{2}} + \delta e^{\nicefrac{-\gamma}{2}})a \d + \delta e^{\nicefrac{-\gamma}{2}}a
\en
one obtains
\be
\gamma = -2\ln \left[ \cos \phi t +i \frac{\omega}{2\phi}\sin \phi t \right]
\en
\be
\beta = -\frac{i\Omega_1\sin \phi t}{\phi \cos \phi t +i \frac{\omega}{2}\sin \phi t}
\en
\be
\delta = -\frac{i\Omega_1^*\sin \phi t}{\phi \cos \phi t +i \frac{\omega}{2}\sin \phi t}\ \ .
\en
Eqn(24), (25) and (26) imply that apart from the physically irrelevant phase factor ($e^{ict}$) the evolution operator for the radiation field is periodic in time, with period, $\tau = \nicefrac{{2\pi}}{\phi}$. If $\tau$ is small compared to the decoherence time scale of the system, this periodicity will be reflected in all the physical properties of the radiation field.

\subsection{Non-classical Radiation}
As discussed in the {\it Introduction}, a state of the radiation field is said to be `non-classical' if the corresponding Glauber-Sudarshan $\mathcal{P}$-function is either negative over some region of the parameter space or is more singular than the Dirac-delta function. With eqn(16)-(17) and (24)-(26) in hand we can calculate the $\mathcal{P}$-function for the time evolved coherent state to check whether it can show non-classical properties or not.

Let initially the field be in the coherent state $|\lambda \rangle$. Therefore,
\be
\rho (t) = U(t)|\lambda \rangle \langle \lambda |U \d (t).
\en
We also have,
\be
\rho (t) = \int d^2 \eta | \eta \rangle \langle \eta | \mathcal{P} (\eta,t)
\en
as the defining relation of the $\mathcal{P}$-function. This implies,
\be
\begin{array}{ccl} \langle -\zeta | \rho (t) | \zeta \rangle & = & \int d^2 \eta \mathcal{P}(\eta , t) \langle - \zeta | \eta \rangle \langle \eta | \zeta \rangle \\
 & = & e^{-|\zeta|^2} \int d^2 \eta \left( \mathcal{P}(\eta , t) e^{-|\eta |^2} \right)e^{\zeta \eta ^* - \zeta ^*\eta } \end{array}
\en
or,
\be
\mathcal{P}(\eta ,t ) = \frac{e^{|\eta|^2}}{\pi ^2} \int d^2 \zeta \langle -\zeta | \rho (t) | \zeta \rangle e^{|\zeta|^2} e^{\zeta ^*\eta - \zeta \eta ^*}.
\en

Using eqn(16),(17) and (24)-(26), one can easily evaluate
\be
\beray
\langle \chi | U(t) | \zeta \rangle & = & e^{\nicefrac{\g}{4}}e^{ict}e^{-\frac{1}{2}\left( |\chi|^2 + |\zeta|^2\right) } \\
& & \times e ^{ \left[\frac{1}{2} \left( \beta \chi ^{*2} + \de \zeta ^2 + 2e^{\nicefrac{\gamma}{2}} \chi ^* \zeta + p \chi ^* + q\zeta + r \right) \right]}
\enray
\en
where
\be
\begin{array}{ccl}
p & = & 2\left( e^{\nicefrac{\g}{2}} - 1 \right) \a + 2\b \a ^*\\
q & = & 2\left( e^{\nicefrac{\g}{2}} - 1 \right) \a ^* + 2\de \a \\
r & = & \b \a ^{*2} + \de \a ^2 + 2\left( e^{\nicefrac{\g}{2}} - 1 \right) |\a |^2.
\end{array}
\en
Therefore,
\be
\begin{array}{ccl}
\langle -\zeta | \rho(t) | \zeta \rangle & = & \langle -\zeta |U(t)|\lambda \rangle \langle \lambda | U(t) | \zeta \rangle \\
 & = & \left| e^{\nicefrac{\g}{4}} \right |^2 e^{-\left( |\zeta |^2 + |\lambda|^2 \right) } e^{ \left[ \frac{1}{2} \left( \beta \zeta ^{*2} + \beta ^* \zeta ^2\right) \right]} \\
 & & \times e^{\left[-\frac{1}{2}  \{ \left( 2 e^{\nicefrac{\g}{2}} \lambda + p\right) \zeta ^* - \left( 2 e^{\nicefrac{\g}{2}} \lambda + p\right) ^* \zeta \} \right]}\\
 & & \times e^{\left[ Re \left( \de \lambda ^2 + q \lambda + r \right) \right]}.
\end{array}
\en
Using eqn(33) in eqn(30)
\be
\mathcal{P} ( \eta ,t) = C \int d^2 \zeta e^{\frac{1}{2} \left( \b \zeta ^{*2} + \b ^* \zeta ^2 \right)} e^{\eta ' \zeta ^* - \eta '^* \zeta}
\en
where
\be
\begin{array}{ccl}
C & = & \frac{e^{|\eta |^2}}{\pi ^2} \left| e^{\nicefrac{\gamma}{4}} \right| ^2 e^{-|\lambda|^2}\exp \left[ Re\left( \de \lambda ^2 + q \lambda + r \right) \right] \\
\eta ' & = & \eta - \frac{\left( 2e^{\nicefrac{\gamma}{2}} \lambda + p \right) }{2}.
\end{array}
\en
Writing $\zeta = \zeta _1 + i \zeta_2$, $\beta = \beta _1 + i\b_2$ and $\eta' = \eta'_1 + i\eta' _2$, where $\zeta _k, \b _k$ and $\eta ' _k\ (k = 1,2)$ are real, one obtains
\be
\mathcal{P} ( \eta ,t) = C \int d\zeta _1 d\zeta _2 e^{\left( \b_1 \zeta_1 ^2 - \b_1 \zeta_2 ^2 + 2\b _2 \zeta _1 \zeta _2 + 2i\zeta _1 \eta _2' - 2i\zeta _2 \eta _1'\right)}
\en
which blows up for all values of $\eta$ (because $\b _1$ is either $> 0$ or $< 0$). Note that $\b$ is a periodic function of time. Also, $\b(t = 0) = 0$ implies that the $\mathcal{P}$-function evaluates to $\delta ^2 (\eta - \lambda)$, as expected for the initial coherent state $|\lambda \rangle $. This check ascertains the correctness of our calculations. 

On the basis of above calculations we can safely conclude that on brief interaction with plasma a coherent state of the radiation field can evolve into a non-classical state. In what follows, we explicitly demonstrate that the field does show the non-classical property of squeezing.

\subsection{Squeezing}\label{4A}
The measure of first order squeezing, as defined in the Appendix A,
\be
\mathcal{D}_1 = \frac{1}{2} \left( \langle a \d a \rangle - |\langle a \rangle |^2 - |\left( \Delta a \right) ^2| \right)
\en
where
\be
\left( \Delta a \right) ^ 2 = \langle a ^2 \rangle - \langle a \rangle ^2.
\en
Now,
\be
\begin{array}{ccl}
\langle a \d a \rangle - |\langle a \rangle |^2 & = & \langle \lambda | U \d (t) a \d a U (t) | \lambda \rangle\\
 & & \quad - \left| \langle \lambda | U \d (t) a U(t) | \lambda \rangle \right| ^2\\
 & = & |g|^2
\end{array}
\en
and
\be
\begin{array}{ccl}
|\left( \Delta a \right) ^2| & = & \left| \langle \lambda | U \d (t) a^2 U(t) | \lambda \rangle - \left( \langle \lambda | U \d (t) a U(t) | \lambda \rangle \right) ^2 \right| \\
 & = & |f||g|.
\end{array}
\en
where
\be
\begin{array}{ccl}
f & = & \cos{\phi t} + \frac{i \omega }{2 \phi } \sin{\phi t} \\
g & = & \frac{i \Omega _1 }{\phi }\sin{\phi t} \qquad  (\text{cf eqn(21)}).
\end{array}
\en
Using eqn(39) and (40) in eqn(37)
\be
\mathcal{D}_1 = \frac{1}{2} |g|\left( |g| - |f| \right).
\en
As mentioned earlier, $|\Omega _1| \ll \omega$. Therefore,
\be
\mathcal{D}_1 \leq 0.
\en
This shows that on short term interaction with plasma the radiation field becomes `non-classical' and shows first order squeezing. Note that $\mathcal{D}_1 (t =0) = 0$, as expected for a coherent state of the radiation field. Also, if the decoherence time of the system is much larger compared to $\nicefrac{2\pi}{\phi}$ (period of $d$ and $e$) the radiation field would return to the state of zero squeezing periodically.
\section{The Thermalized Radiation Field}
On long term interaction (interaction time greater than the decoherence time scale of the system) the field comes to thermal equilibrium with the plasma. Considering the entire system to be in contact with a heat bath at constant temperature $T$, the time independent density matrix for the radiation field is given by
\be
\rho = \frac{e^{-\nicefrac{H}{\theta}}}{Z}
\en
where $\theta = k_B T$ and
\be
Z = Tr \left( e^{-\nicefrac{H}{\theta}} \right).
\en

We now exploit the connection between the unnormalized density matrix ($e^{-\nicefrac{H}{\theta}}$) and the time evolution operator U(t). The former is obtainable from the later by analytically continuing the time parameter to the purely imaginary value of $-i\theta ^{-1}$. Thus from eqn(16) the unnormalized density matrix
\be
e^{-\nicefrac{H}{\theta}} = e^{\nicefrac{c}{\theta}} D \d (\alpha ) e^{-\nicefrac{P}{\theta}}D (\alpha).
\en
Also, following section \ref{dis}, we can immediately write down the disentangled form for the SU(1,1) operator in eqn(46), where the disentanglement coefficients are now given (by replacing $t$ with $-i\theta ^{-1}$) by
\be
\gamma = -2\ln \left[ \cosh{\frac{\phi}{\theta}} + \frac{\omega}{2\phi}\sinh{\frac{\phi}{\theta}} \right]
\en
\be
\beta = -\frac{\Omega_1\sinh{\frac{\phi}{\theta}}}{\phi \cosh{\frac{\phi}{\theta}} + \frac{\omega}{2}\sinh{\frac{\phi}{\theta}}}
\en
\be
\delta = -\frac{\Omega_1^*\sinh{\frac{\phi}{\theta}}}{\phi \cosh{\frac{\phi}{\theta}} + \frac{\omega}{2}\sinh{\frac{\phi}{\theta}}} = \b ^*.
\en

It is now a matter of algebra to check the existence of the $\mathcal{P}$-function. From eqn(30), we know that the $\mathcal{P}$-function is the two dimensional Fourier transform of $\langle -\zeta | \rho | \zeta \rangle e^{|\zeta|^2}$. If we can show that this is a well behaved function of $\zeta$, the existence of the corresponding $\mathcal{P}$-function would be ascertained. If, however, $\langle -\zeta | \rho | \zeta \rangle e^{|\zeta|^2}$ is not well behaved for some range of $\theta$ we can expect to observe non-classical properties of the radiation field in that temperature range.
\be
\beray
Z \langle -\zeta | \rho | \zeta \rangle & = & e^{\nicefrac{c}{\theta}} \langle -\zeta | D \d (\alpha ) e^{-\nicefrac{P}{\theta}}D (\alpha) | \zeta \rangle \\
 & & \\
 & = & e^{\nicefrac{c}{\theta} + \nicefrac{\g}{4}} e^{ \left[ - \frac{1}{2} \{ 2\left( 1 + e^{\nicefrac{\g}{2}}\right) |\zeta |^2 - \b \zeta ^{*2} - \de \zeta ^2 \} \right] } \\
 & & \times e^{ \left[ \{ \a \left( 1 - e^{\nicefrac{\g}{2}} \right) - \b \a ^* \}\zeta ^* - \{ \a ^*\left( 1 - e^{\nicefrac{\g}{2}} \right) - \de \a \}\zeta \right] } \\
 & & \times e^{ \left[ -\frac{1}{2} \{ 2\left( 1 - e^{\nicefrac{\g}{2}}\right) |\a |^2 - \b \a ^{*2} - \de \a ^2 \}\right] }
\enray
\en
where we used the disentangled form of $e^{-\nicefrac{P}{\theta}}$. Now, writing $\zeta$ as $\zeta _1 +i \zeta _2$, where $\zeta _1 $ and $\zeta _2$ are real variables, one finds that $\langle -\zeta | \rho | \zeta \rangle e^{|\zeta|^2}$ is a Gaussian with the coefficient matrix for the quadratic part being
\be
\mathcal{M} = \left(
\begin{matrix}
e^{\nicefrac{\g}{2}} - \nicefrac{\b}{2} - \nicefrac{\de}{2} & - Im(\b) \\
-Im(\b) & e^{\nicefrac{\g}{2}} + \nicefrac{\b}{2} + \nicefrac{\de}{2}
\end{matrix} \right).
\en
Clearly, $Tr(\mathcal{M}) > 0$, and
\be
\beray
Det(\mathcal{M}) & = & e^{\g} - | \b |^2\\
 & = & e^{\g} \left( 1 - \frac{|\Omega_1|^2}{\phi ^2} \sinh ^2{ \frac{\phi}{\theta} } \right).
\enray
\en
Of course, if $\Omega _1 = 0$, this is always positive. However, though $|\Omega _1|$ is very small (compared to $\omega$) in systems of physical interest, it is easy to see that for temperatures smaller than
\be
\omega / \left[ 2 \ln \left( \omega / \Omega _1\right)\right]\equiv \theta _c,
\en
$Det(\mathcal{M})$ will become negative. At these temperatures, the Gaussian form will have a growing exponent, and thus the $\mathcal{P}$-function would cease to exist. Thus, in contrast to the thermalized free field (cf $\Omega_1 = 0$ case), we see that the thermalized radiation field in laser plasma interaction can become non-classical at sufficiently low temperatures. In the following subsection we explicitly show that the thermalized radiation field in this case does exhibit the non-classical property of squeezing in the above predicted temperature range.

\subsection{Squeezing}\label{5A}
We will use the disentanglement relation first to calculate the partition function in the presence of `sources'. That is, we will replace the $e^{-\nicefrac{P}{\theta}}$ in eqn(46) by $e^{\varepsilon ( a \d - \a ^*)}e^{-\nicefrac{P}{\theta}}e^{\eta ( a - \a )}$. Note that we have not introduced the source terms as addenda to the Hamiltonian, but rather have put them in as disentangled terms. This has the advantage of simplifying the relationship between the partition function and various mean values that we need to calculate to evaluate $\mathcal{D}_1$ and $\mathcal{D}_2$ eqn(A8). Thus, with the unnormalized density matrix with sources denoted by $\tilde{\rho} ( \varepsilon , \eta ) $, we find
\be
\begin{array}{ccl}
Z (\varepsilon , \eta) & = & Tr \left[ \tilde{\rho} (\varepsilon, \eta) \right]\\
 & = & Tr \left[ e^{\nicefrac{c}{\theta}} D \d (\alpha ) e^{\varepsilon ( a \d - \a ^*)}e^{-\nicefrac{P}{\theta}}e^{\eta ( a - \a )}D (\alpha) \right].
\end{array}
\en
This implies
\be
\langle a^n a ^{\dagger m}\rangle = \frac{1}{Z} \frac{\partial ^{m+n}}{\partial ^m \varepsilon \partial ^n \eta} Z(\varepsilon , \eta) \vline \begin{array}{l}
\\
_{\varepsilon = \eta = 0}.
\end{array}
\en
Shifting the relative position of the source operators with respect to the factor $e^{-\nicefrac{P}{\theta}}$ will give us a similar formula for the mean values of normally ordered operator products. However, we will soon see that the advantages we gain in using the coherent state basis in calculating the partition function makes the form used most convenient.

Using the coherent state basis
\be
Z (\varepsilon , \eta) = \frac{e^{\nicefrac{c}{\theta}}}{\pi} \int d^2 \zeta \langle \zeta | e^{\varepsilon ( a \d - \a ^*)}e^{-\nicefrac{P}{\theta}}e^{\eta ( a - \a )} | \zeta \rangle.
\en
Where we used the cyclic property of $Tr()$. Now, using the disentangled form of $e^{-\nicefrac{P}{\theta}}$
\be
\beray
Z (\varepsilon , \eta) & = & \frac{e^{\nicefrac{c}{\theta} + \nicefrac{\gamma}{4}}}{\sqrt{\left( 1 - e^{\nicefrac{\g}{2}}\right)^2  - \b \de}} \\
& & \times \exp\left[ \frac{2\left( 1 - e^{\nicefrac{\g}{2}}\right) \varepsilon \eta + \de \varepsilon ^2 + \b \eta ^2}{2\{ \left( 1 - e^{\nicefrac{\g}{2}}\right) ^2 - \b \de \} } - \varepsilon \a ^* - \eta \a \right].
\enray
\en
This is a simple Gaussian in $\eta$ and $\varepsilon$. This allows us to calculate the various expectation values simply in terms of the parameters $\b$, $\g$ and $\de$. Denoting the derivatives of the logarithm of the partition function by $t_\varepsilon$ $\left( \text{for }\left( \frac{\partial \ln Z}{\partial \varepsilon} \right) _{\varepsilon = \eta = 0}\right) $, $t_{\varepsilon \varepsilon \eta}$ $\left( \text{for } \left( \frac{\partial^3 \ln Z}{\partial^2 \varepsilon \partial \eta }\right) _{\varepsilon = \eta = 0} \right)$, etc., we find that the only non-zero $t$'s are
\be
\beray
t_\eta & = & t_\varepsilon ^* = -\a \\
t_{\eta \eta} & = & t_{\varepsilon \varepsilon}^* = - \frac{\Omega _1}{2 \phi} \coth{\frac{\phi}{2 \theta}} \\
t _{\varepsilon \eta} & = &\frac{1}{2} \left( 1 + \frac{\omega}{2\phi}\coth{\frac{\phi}{2 \theta}}\right).
\enray
\en

One can now easily show that the following relations hold
\be
\beray
\langle a \rangle & = & t _\eta\\
\langle a ^2 \rangle & = & t_{\eta \eta} + t _\eta ^2 \\
\langle a a \d \rangle & = & t _{\varepsilon \eta} + t_\varepsilon t_\eta \\
\langle a^2 a ^{\dagger 2} \rangle & = & t _{\eta \eta} t_{\varepsilon \varepsilon} + 2 t_{\varepsilon \eta} ^2 + 4t_\varepsilon t_\eta t_{\varepsilon \eta} \\
& & + t_\eta ^2 t_{\varepsilon \varepsilon} + t_\varepsilon ^2 t_{\eta \eta} + t_\eta ^2 t_\varepsilon ^2\\
\langle a ^4 \rangle & = & t _{\eta} ^4 + 6 t _{\eta} ^2 t _{\eta \eta} + 3 t _{\eta \eta}^2 \ .
\enray
\en
Using eqn(37)
\be
\beray
\mathcal{D}_1 & = & \frac{1}{2}\left( t_{\varepsilon \eta} - | t _{\eta \eta}| - 1\right)\\
 & = & \frac{1}{2}\left[ \left(\frac{\omega}{4\phi} - \frac{|\Omega _1|}{2\phi}\right) \coth {\frac{\phi}{2\theta}} - \frac{1}{2}\right].
\enray
\en
Since, for physical situations $|\Omega _1| \ll \omega$,
\be
\mathcal{D}_1 \simeq \frac{1}{4} \left[ \left(1 - \frac{2|\Omega _1|}{\omega}\right) \coth {\frac{\omega}{4\theta}} - 1\right].
\en
Clearly, $\mathcal{D}_1$ goes negative for $\theta < \theta _c$ (eqn(53)). This is in agreement with the temperature range where the $\mathcal{P}$-function is not defined.

We note here, that if we follow Zhang et al's \cite{Zhang} definition of squeezing parameter
\be
\beray
\mathcal{D}_1 ^{\text{Zhang}}& = & \frac{1}{4} \left[ 2 \langle a \d a\rangle + \langle a ^2\rangle + \langle a ^{\dagger 2}\rangle - \left( \langle a \rangle + \langle a \d \rangle \right) ^2\right]\\
 & = & \frac{1}{2} \left[ t _{\varepsilon \eta} + Re(t _{\eta \eta}) - 1\right].
\enray
\en
For the physical situation analysed by Ben-Aryeh and Mann \cite{Ben}, $\Omega _1$ being a real positive quantity and $\Omega_1 \ll \omega$,
\be
\mathcal{D}_1 ^{\text{Zhang}} = \frac{1}{2}\left[ \left(\frac{\omega}{4\phi} - \frac{\Omega _1}{2\phi}\right) \coth {\frac{\phi}{2\theta}} - \frac{1}{2}\right].
\en
Which would predict squeezing in the same temperature range as predicted by the measure of squeezing used by us. But, for situations where $Re(\Omega _1) < 0$,
\be
\mathcal{D}_1 ^{\text{Zhang}}\geq \frac{1}{4} \left[ \coth{\frac{\phi}{2 \theta}} - 1\right] \geq 0.
\en
That is, Zhang et al's definition of the first order squeezing parameter does not predict squeezing in such situations. Hence, in general the definition of first order squeezing used by us is more sensitive than the one proposed by Zhang et al. In Appendix A (eqn(A9)) we argue that the $k^{\text{th}}$ order squeezing parameter defined there (which we have used in our calculations) is a better measure of $k^{\text{th}}$ order squeezing than the one proposed by Zhang et al;
\be
\mathcal{D}_k \leq \mathcal{D}_k ^{\text{Zhang}}.
\en

Using eqn(65), one can make an useful comment about the statistics of photon number in the thermalized radiation field. Firstly we note,
\be
\beray
\left( \Delta n \right) ^2 - \langle n\rangle & = & |t_{\varepsilon \varepsilon} | ^2 + \left( t_{\varepsilon \eta} - 1\right) ^2 \\
 & & + 2|\a |^2 \left( t_{\varepsilon \eta} - 1\right) + \left( t _{\varepsilon \varepsilon} \a ^2 + t_{\eta \eta} \a ^{*2}\right) \\
 & = & \mathcal{D}_2 ^{\text{Zhang}}
\enray
\en
where $n = a \d a$ and $\Omega _1$ is taken to be real to arrive at the last equality. Next,
\be
\begin{array}{c}
\mathcal{D}_2 =\\
\left[ t _{\varepsilon \eta} ^2 + 2|\a |^2 t _{\varepsilon \eta} - 2| \a | ^2 + 1 - 2t _{\varepsilon \eta} - |2\a ^2 t _{\eta \eta} + t _{\eta \eta} ^2 |\right].
\end{array}
\en
Note that $t_{\eta \eta}$ is real. Therefore,
\begin{widetext}
\be
\begin{array}{c}
\mathcal{D}_2 \geq \\
\left[ t _{\varepsilon \eta} ^2 + 2|\a |^2 t _{\varepsilon \eta} - 2| \a | ^2 + 1 - 2t _{\varepsilon \eta} - 2|\a |^2 t _{\eta \eta} - t _{\eta \eta} ^2 \right] =\\
\left[ \left( t _{\varepsilon \eta} + t_{\eta \eta}\right) \left( t _{\varepsilon \eta} - t_{\eta \eta}\right) + 2|\a |^2 \left( t _{\varepsilon \eta} - t_{\eta \eta}\right) - 2|\a |^2 -2 t_{\varepsilon \eta} +1 \right].
\end{array}
\en
\end{widetext}
Revoking $\Omega _1 \ll \omega$
\be
\begin{array}{c}
\mathcal{D}_2 \gtrsim \\
\left[ t _{\varepsilon \eta} ^2 + 2|\a |^2 t _{\varepsilon \eta} - 2| \a |^2 - 2 t _{\varepsilon \eta} +1 \right] = \\
\left[ \left( t_{\varepsilon \eta} - 1\right)^2 + 2|\a |^2 \left( t_{\varepsilon \eta} - 1\right) \right] > 0
\end{array}
\en
because $t_{\varepsilon \eta} > 1$. From eqn(65), (66) and (69)
\be
\left( \Delta n \right) ^2 - \langle n\rangle > 0.
\en
Which implies that the photon number in the thermalized radiation field follows super-Poissonian statistics; a property usually exhibited by `classical' radiation fields.

\section{Conclusion}
On the basis of this work we arrive at the following important conclusions. Firstly, we found via explicit calculations that unitary time evolution of the state of the radiation field under the Hamiltonian in eqn(5) would lead to the production of non-classical states having indefinite $\mathcal{P}$-functions. Among the observable properties, these non-classical field states would exhibit first order squeezing (as also found by Ben-Aryeh and Mann \cite{Ben}), with the amount of squeezing depending periodically on the interaction time between the field and the plasma.

Secondly, we could also show that even the thermalized radiation field state will have an indefinite $\mathcal{P}$-function below a certain temperature ($\theta _c$). Also, using a new measure of $k^{\text{th}}$ squeezing, expounded up on in Appendix A, the thermalized radiation field is shown to exhibit first order squeezing at temperatures below $\theta _c$. The authors feel that it would be interesting to take a closer look at exactly what is happening at temperatures close to $\theta _c$, so that, contrary to the belief, even thermalized field starts exhibiting non-classical properties.

Here we must point out that for the case $Re(\Omega_1) > 0$ in eqn(5), the measure of $k^{\text{th}}$ order squeezing proposed by Zhang et al \cite{Zhang} does also predict first order squeezing in the considered thermalized field, for temperatures below $\theta _c$. But, it does not predict squeezing in either first or second order for $Re(\Omega _1)<0$ case.

Finally, using eqn(A9), we establish that the considered thermalized radiation field would follow super-Poissonian statistics at all temperatures.

\begin{acknowledgments}
This work is an extension over a part of ADG's PhD thesis \cite{ADG1}. After proof reading by PG and AVM several results, including the definition of the new measure of $k^{\text{th}}$ order squeezing described and used above, and some text have been borrowed from \cite{ADG1} with ADG's permission.
\end{acknowledgments}

\appendix
\section{$k^{\text{th}}$ order amplitude squeezing}
We define $k^{\text{th}}$ order quadrature operators as follows,
\be
X_k(\theta ) = \frac{1}{2} \left( a^k e ^{-i \theta} + a^{\dagger k} e ^{i \theta}\right).
\en
The $k^{\text{th}}$ order quadrature operators used by Zhang et al \cite{Zhang} to define $k^{\text{th}}$ order squeezing are $X_k (0)$ and $X_k (\nicefrac{\pi}{2})$. More generally, one can use $X_k (\theta )$ and $X_k (\theta + \nicefrac{\pi}{2} )$ to define $k^{\text{th}}$ order squeezing.
\be
\left[ X_k (\theta) , X_k (\theta + \nicefrac{\pi}{2} ) \right] = \frac{i}{2} \left[ a ^k , a ^{\dagger k}\right]
\en
which implies
\be
\left( \Delta X _k (\theta) \right) ^2 \left( \Delta X _k (\theta + \nicefrac{\pi}{2}) \right)^2 \geq \frac{1}{16} | \langle \left[ a ^k , a ^{\dagger k} \right] \rangle | ^2.
\en
The quadrature $X_k(\theta )$ is said to exhibit $k^{\text{th}}$ order squeezing if
\be
\left( \Delta X _k (\theta) \right) ^2 < \frac{1}{4} | \langle \left[ a^k, a^{\dagger k}\right] \rangle |.
\en
Let
\be
\mathcal{D}_k (\theta ) \equiv \left( \Delta X _k (\theta) \right) ^2 - \frac{1}{4} | \langle \left[ a^k, a^{\dagger k}\right] \rangle |
\en
or,
\begin{widetext}
\be
\mathcal{D}_k (\theta ) = \left\{ 
\beray
\frac{1}{2} \left[ \langle a^{\dagger k} a ^k \rangle - |\langle a ^k \rangle |^2 + |\Delta a ^k|^2 \cos{2(\phi - \theta) } \right] \quad \mbox{if $[a^k, a^{\dagger k} ] > 0$}\\
\\
\frac{1}{2} \left[ \langle  a ^k a^{\dagger k} \rangle - |\langle a ^k \rangle |^2 + |\Delta a ^k|^2 \cos{2(\phi - \theta) } \right] \quad \mbox{if $[a^k, a^{\dagger k} ] < 0$}
\enray \right.
\en
where $\phi = Arg(\Delta a ^k )$. Note, Zhang et al \cite{Zhang} did not consider the case of $[a^k, a^{\dagger k} ] < 0$. Incorporating this case in their definition of squeezing parameter,
\be
\mathcal{D}_k ^{\text{Zhang}} = \mathcal{D}_k (\theta = 0 ) = \left\{ 
\beray
\frac{1}{2} \left[ \langle a^{\dagger k} a ^k \rangle - |\langle a ^k \rangle |^2 + |\Delta a ^k|^2 \cos{2\phi } \right] \quad \mbox{if $[a^k, a^{\dagger k} ] > 0$}\\
\\
\frac{1}{2} \left[ \langle  a ^k a^{\dagger k} \rangle - |\langle a ^k \rangle |^2 + |\Delta a ^k|^2 \cos{2\phi } \right] \quad \mbox{if $[a^k, a^{\dagger k} ] < 0$.}
\enray \right.
\en
According to Zhang et al \cite{Zhang}, a field exhibits $k^{\text{th}}$ order squeezing if $\mathcal{D}_k ^{\text{Zhang}} < 0$.

Using the liberty of choosing $\theta$ in eqn(A4) we arrive at a stricter definition of squeezing. We will say that a field is squeezed to order k if $\mathcal{D}_k < 0$, where
\be \label{A8}
\mathcal{D}_k \equiv \left\{ 
\beray
\frac{1}{2} \left[ \langle a^{\dagger k} a ^k \rangle - |\langle a ^k \rangle |^2 - |\Delta a ^k|^2 \right] \quad \mbox{if $[a^k, a^{\dagger k} ] > 0$}\\
\\
\frac{1}{2} \left[ \langle  a ^k a^{\dagger k} \rangle - |\langle a ^k \rangle |^2 - |\Delta a ^k|^2 \right] \quad \mbox{if $[a^k, a^{\dagger k} ] < 0$}
\enray \right.
\en
Eqn(A8) defines our $k^{\text{th}}$ order squeezing parameter ($\mathcal{D}_k$), which has been used as the measure of squeezing in this paper. Clearly,
\be
\mathcal{D}_k \leq \mathcal{D}_k ^{\text{Zhang}}.
\en
\end{widetext}

\bibliographystyle{unsrt}
\bibliography{report}

\end{document}